\documentclass[prl,twocolumn,superscriptaddress,showpacs]{revtex4}
\usepackage{epsfig,amsmath,amssymb,graphicx,color,calc}            

\newcommand{\be}{\begin{equation}}
\newcommand{\ee}{\end{equation}}
\newcommand{\ba}{\begin{eqnarray}}
\newcommand{\ea}{\end{eqnarray}}

\begin{document}

\title{Increasing the density melts ultrasoft colloidal glasses}

\author{Ludovic Berthier}
\affiliation{Laboratoire des Collo{\"\i}des, Verres
et Nanomat{\'e}riaux, UMR CNRS 5587, Universit{\'e} Montpellier 2,
34095 Montpellier, France}

\author{Angel J. Moreno}
\affiliation{Centro de F\'{\i}sica de Materials (CSIC, UPV/EHU) and Materials
Physics Center MPC, Paseo Manuel de Lardizabal 5, E-20018 
San Sebasti\'an, Spain}

\author{Grzegorz Szamel}
\affiliation{Department of Chemistry, Colorado State University, 
Fort Collins, CO 80523}

\date{\today}
 
\begin{abstract}
We use theory and simulations to investigate the existence of amorphous glassy 
states in ultrasoft colloids.
We combine the hyper-netted chain approximation with mode-coupling 
theory to study the dynamic phase diagram of soft repulsive spheres 
interacting with a Hertzian potential, focusing on low temperatures and large
densities. At constant temperature, we find that an amorphous glassy state
is entered upon compression, as in colloidal hard spheres, but the glass
unexpectedly melts when density increases further. We attribute this 
re-entrant fluid-glass transition to particle softness, and 
correlate this behaviour to previously reported anomalies in 
soft systems, thus emphasizing its generality.
The predicted fluid-glass-fluid sequence is confirmed numerically.
\end{abstract}

\pacs{05.20.Jj, 64.70.qd}


\maketitle

The hard sphere system is an emblematic model in statistical 
physics. It is a useful theoretical model
for liquids which can be studied experimentally using 
colloids~\cite{pusey}, its density driven fluid-crystal and fluid-glass 
transitions have been extensively studied. Recently, there has 
been growing interest in  systems composed of ultrasoft 
particles, which can be 
realized experimentally using colloidal particles made of 
star-shaped polymers or microgels~\cite{likosreview}. Since softness
introduces (at least) one more dimension to the phase diagram,  
new physical behaviour can be expected from the competition between 
entropy, which governs the behaviour of hard spheres, and
energy, which appears when soft particles overlap.

As a result, the physics of systems consisting of soft particles is
considerably richer than that of hard ones. 
Unusual equilibrium phase diagrams arise depending on the
particular features of particle softness. These include 
cluster crystal structures at
arbitrary large densities~\cite{mladek} and a 
non-monotonic density dependence of the freezing 
transition~\cite{stillinger,langlikos,ardlouis}
with a complex cascade of 
crystalline states~\cite{frenkel}. 
In the latter case, the thermodynamically stable fluid phase 
also exhibits density anomalies, at the 
structural~\cite{ardlouis,langlikos,frenkel,jacquin} and 
dynamical~\cite{krekelbergtruskett,frenkel,egorov} levels.

Exploring the existence of amorphous solid states at low
temperature and large density in soft particle materials is an emerging
field. The glassy dynamics of star polymers have been
investigated both theoretically and experimentally, and, in particular,  
non-trivial phase diagrams and novel glassy states have 
been described in star-polymer 
mixtures~\cite{likosreview,foffi,mayerzaccarelli}. 
Several studies were dedicated to the perturbations brought 
about by adding an energy scale to the hard sphere glassy behaviour.
Such a perturbation not only shifts the glass transition to larger 
densities~\cite{tom,harmonic}
but also affects qualitatively the dynamics near the transition, 
in particular the glass fragility~\cite{tom,dave,ken}.
Experiments have also investigated the physics of glass aging~\cite{aging},
mostly in microgels, with a particular 
interest on the jamming transition 
intervening deep into the glass phase~\cite{zhang,luca}.

In this work we investigate whether the anomalous 
equilibrium phase diagram and transport properties 
of soft repulsive sphere systems also impact 
their transition to a non-ergodic glassy state.
This possibility was evoked for star polymers~\cite{foffi}
for a narrow range of star functionalities, but was not 
confirmed numerically because
crystallization intervened before slow dynamics set in. 
Other theoretical studies of soft repulsive spheres focused on 
densities close to the hard sphere transition, leaving 
the behaviour at large densities unexplored~\cite{harmonic}. 

Our main finding is that an `anomalous' density dependence
of the glass transition temperature is found both theoretically
and in computer simulations for the well-studied Hertzian
model for ultrasoft particles. Thus, it is possible to melt a glass made
of soft particles by 
increasing its density, a scenario that has not been observed 
yet in experiments. This is surprising on general 
grounds, because particle 
crowding typically yields jammed, rather than fluid, states for 
`usual' particles such as hard spheres or molecular fluids.
Indeed, traffic jams are rarely eased by putting more cars
on the road.    

We consider $N$ particles in a volume $V$ 
interacting with 
a soft, repulsive, pairwise additive Hertzian potential,  
\begin{equation}
V(r < \sigma) = \epsilon (1-r/\sigma)^{5/2},
\label{pair}
\end{equation}
and $V(r \ge \sigma)=0$. 
Originally derived to describe
the elastic repulsion between deformable spheres, 
the Hertzian potential has recently 
been used to study theoretically the physics of 
soft repulsive particles in various contexts, and seems to 
describe well the interaction between microgel colloids~\cite{zhang}.
Its physics is very similar to the harmonic sphere model 
inspired by the physics of foams~\cite{durian}, 
or to the Gaussian core model~\cite{stillinger},
derived from polymer physics. 
The control parameters
of the system are the particle number density, $\rho=N\sigma^3/V$,  
and the ratio of the temperature $T$ 
and energy scale $\epsilon$. Since particles can overlap 
at a finite energy cost, $\rho$ can significantly 
exceed 1 and, indeed, we are interested in this 
high density regime. 
We use reduced units, measuring
lengths in units of $\sigma$, and temperatures in units of $\epsilon$.

We study the structure and dynamics of Hertzian spheres combining theory and
simulations. For the theory, we closely follow an earlier 
investigation of harmonic spheres to which we refer for further 
details and equations~\cite{harmonic}. We examine a monodisperse 
system of Hertzian spheres using the hyper-netted chain (HNC) 
approximation~\cite{hansen} to obtain the pair correlation 
function $g(r)$ and the
structure factor $S(q)$ for a given state point $(\rho,T)$.
We then inject $S(q)$ in the dynamical equations of motion
for time correlation functions derived within 
mode-coupling theory (MCT)~\cite{gotze}, which we solve 
numerically~\cite{elijah}.
We obtain the time dependence of density-density correlation functions for a 
wide range of wave-vectors for a broad range of densities
and temperatures.  

To verify theoretical predictions we perform molecular 
dynamics (MD) simulations of a polydisperse Hertzian sphere system.
Polydispersity is needed to prevent crystallization or fractionation
that easily occur in the monodisperse system.
The particle diameters $\sigma_i$ 
are randomly drawn from a flat probability 
distribution with $0.826 \leq \sigma_i \leq 1.126$. 
Two particles $i,j$  interact through Eq.~(\ref{pair}) with 
$\sigma = (\sigma_i + \sigma_j)/2$. We use $N = 9000$ and 
periodic boundary conditions.
Equilibration 
is achieved through periodic
velocity rescaling, production runs are carried 
out in the microcanonical ensemble. 

\begin{figure}
\includegraphics[width=8.5cm]{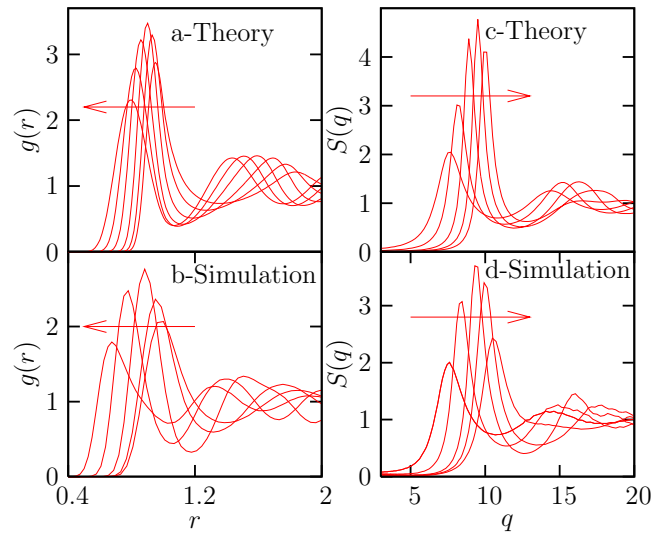}
\caption{\label{4structure} Nonmonotonic evolution of structural
quantities with increasing density (arrows) as predicted by theory (top)
and obtained from MD simulations (bottom). 
a) Predicted pair correlation function for
$T=$ 0.00287 and $\rho=1.1$, 1.4, 1.7, 2.1, 2.43, 2.84; 
b) Numerically measured $g(r)$ for 
$T=$ 0.0043 and $\rho=$ 1.2, 1.45, 2.05, 3.0, 4.05;
c) Predicted structure factor for
$T=$ 0.00287 and $\rho=1.1$, 1.5, 2.1, 2.63, 2.84;
d) Numerically measured $S(q)$ for 
$T=$ 0.0043 and $\rho=$ 1.2, 1.7, 2.4, 3.0, 4.05.}
\end{figure}

We now discuss two representative
quantities characterizing the structure of the fluid at various state points.
In Fig.~\ref{4structure}-a, we show the evolution with the density 
of the pair correlation
function $g(r)$ at constant temperature, $T=0.00287$, as predicted 
by the HNC approximation. As expected,
with increasing density particles are getting closer to each other
and as a result, the position of the first peak of $g(r)$ 
moves towards smaller values of $r$. 
The monotonic shift of the peak position of $g(r)$ is accompanied by an
`anomalous', non-monotonic evolution of the peak height~\cite{frenkel}. 
Specifically, with increasing density from 
$\rho=1.1$ to 1.7 the height of the first peak increases, 
but it decreases when increasing density further from
$\rho=1.7$ to 2.84. This evolution is `anomalous' in the sense
that the first peak of $g(r)$ typically increases monotonically 
with density in simple liquids at constant temperature.  
The locus of the density anomaly evolves with temperature, 
as studied in detail for the harmonic potential~\cite{jacquin}. 

Since $g(r)$ and the structure factor $S(q)$ are related 
via Fourier transform~\cite{hansen}, 
it is not surprising that the same anomalous 
trend is observed in Fig.~\ref{4structure}-c, which 
shows the HNC-derived $S(q)$ for the same temperature.
With increasing density
the position of the first 
peak shifts to larger $q$, corresponding to smaller
interparticle distances. Simultaneously, 
the peak height exhibits a non-monotonic variation.
The maxima of $g(r)$ and $S(q)$ typically occur 
at slightly different densities, respectively 
$\rho \approx 1.7$ and $\rho \approx 2.6$
at $T=0.00287$.

The trends exhibited by $g(r)$ and $S(q)$ calculated using the 
HNC approximation are also evident in the simulations. 
In Fig.~\ref{4structure}-b,d we show the 
pair correlation function and the structure factor
obtained in simulations at constant $T=0.0043$ and various densities.
Both quantities exhibit a non-monotonic evolution with density, 
$g(r)$ having its maximum near $\rho \approx 2.05$, and $S(q)$ 
near $\rho \approx 2.4$. Thus, 
the two maxima are closer to each other
in simulations than according to theory. 
However the similarity between theory and simulations is quite 
striking from these plots, 
showing that HNC is performing reasonably well for the explored
state points. 
Quantitative agreement cannot be expected
as we are using different particle size distributions in both 
approaches. In particular, we believe that polydispersity 
is responsible for the fact that the first peaks of both $g(r)$ and 
$S(q)$ obtained from MD are quite a bit smaller and broader than 
those calculated using the HNC approximation.

The non-monotonic dependence of the height of the
first peak of $g(r)$ on density was previously discussed for several similar
models. Physically, it 
originates from a competition between energy and entropy with
an `anomalous' outcome due to the extreme softness of the 
particles~\cite{langlikos,ardlouis,frenkel,jacquin}.
Briefly, at smaller densities, the system can easily minimize its free energy 
by avoiding the energetically costly particle overlaps. 
In this regime, the system 
effectively behaves as hard spheres, with 
$g(r)$ and $S(q)$ becoming sharper as $\rho$ increases.  
At larger densities, it becomes difficult, 
and thus entropically costly,  
for the system to find configurations with no overlap  
(note that the nominal volume fraction $\phi=\pi\rho/6$ may exceed 100\%).
Rather than trying to do the impossible, particles increase 
their overlaps, have effectively more space, and may thus access 
a larger number of configurations.
This increasing disorder leads to the broadening 
of the first peak of correlation functions.
The entropy gain can be larger than the energy cost
of the overlaps because the Hertzian potential is bounded 
at any separation.

The above described competition between entropic and energetics effects
is purely statistical. It is reasonable to expect that the resulting 
structural anomalies will cause a non-trivial density dependence 
of the equilibrium 
freezing transition~\cite{frenkel}.
However, this competition and the anomalies are not obviously relevant 
to the nonequilibrium glass transition, which is related to the increasing
difficulty to transport matter with increasing particle crowding. 
Our main result is that the anomalous density dependence of 
equilibrium structural quantities is indeed accompanied
by the non-monotonic dependence of 
structural relaxation and ergodicity breaking. 
Anomalous density dependence of transport properties 
were found before in the thermodynamically stable, non-glassy soft particle
fluid~\cite{langlikos,krekelbergtruskett,frenkel,egorov,foffi}, 
but remained essentially unexplored for the deeply supercooled state 
and for the glass. 

\begin{figure}
\includegraphics[width=8.5cm]{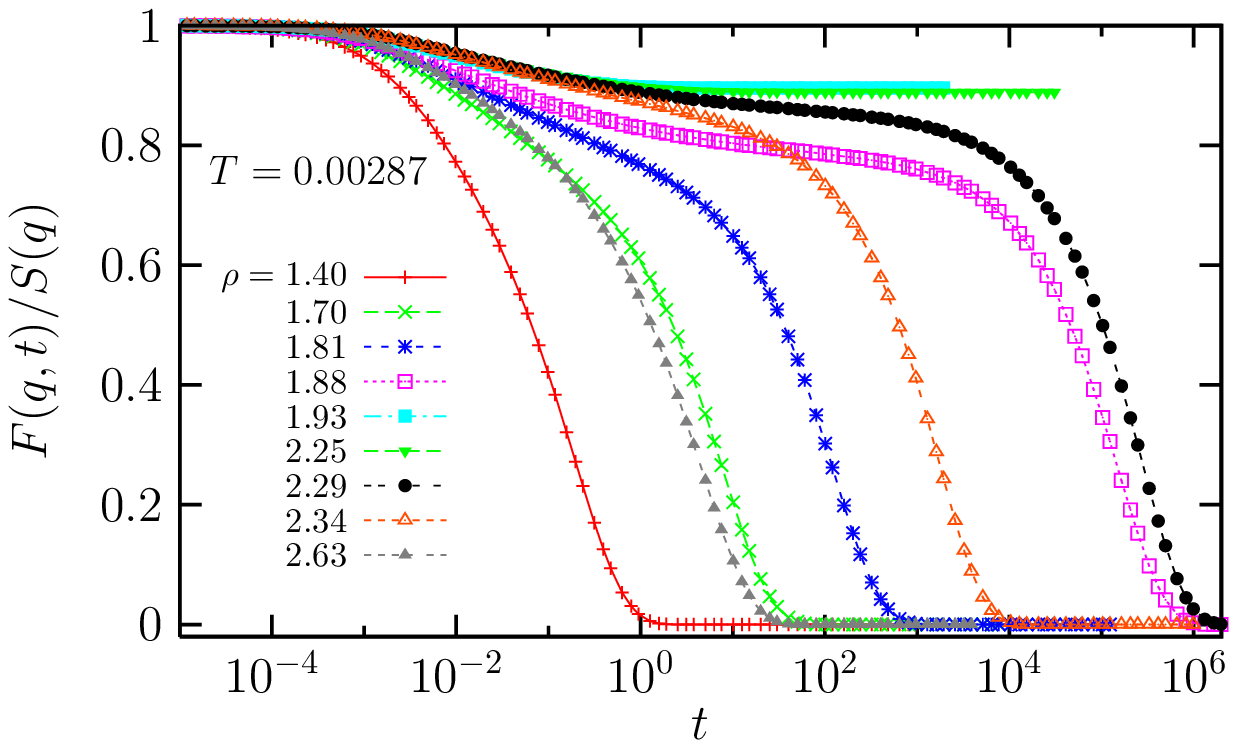}
\includegraphics[width=8.5cm]{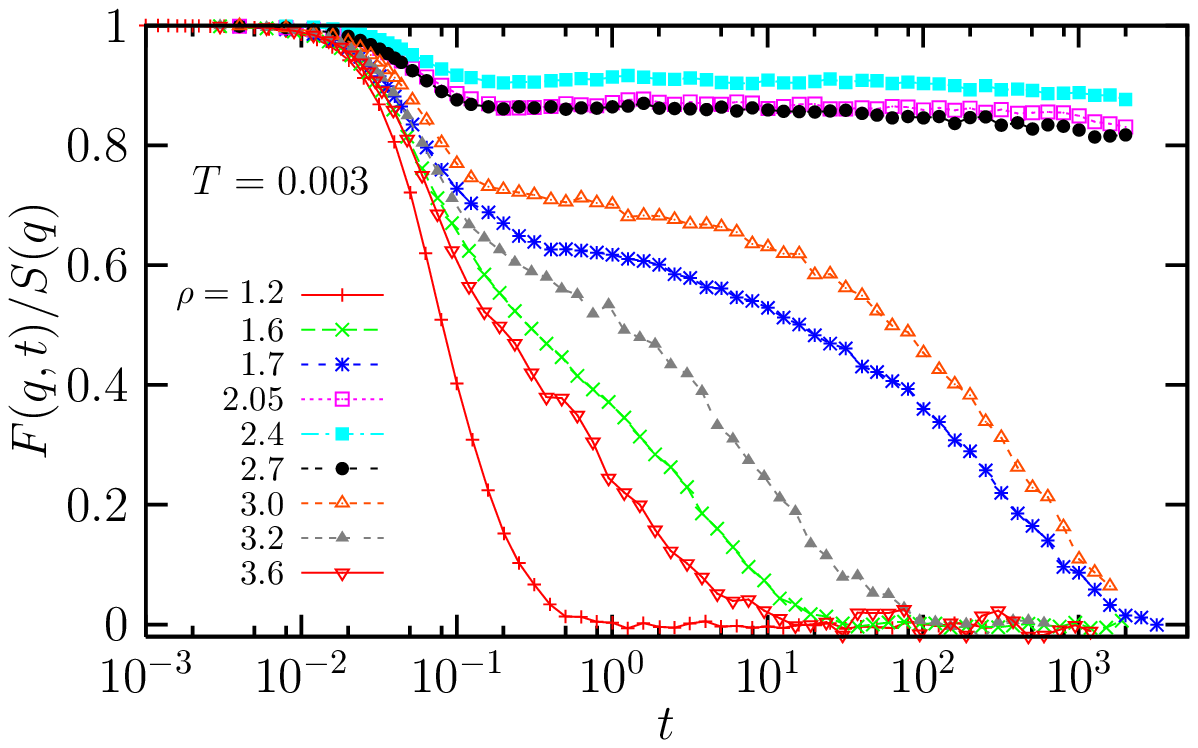}
\caption{\label{FqtMCT} Nonmonotonic evolution of time correlation functions 
with density at constant temperature 
for theory (top, $q=9.1$) and simulations (bottom, $q=9.4$). 
A two-step decay initially develops when $\rho$ increases, 
non-ergodic glass states are obtained at intermediate densities, 
and the glass melts upon further compression.}
\end{figure}
 
In Fig.~\ref{FqtMCT} we show the $\rho$ dependence of the 
density-density time correlators 
as obtained from both theory and simulations. 
As for structural quantities we show the correlators 
at constant $T$ and increasing $\rho$, which exposes
the most remarkable features of the data, but a much
broader range of state points was explored.
Figures~\ref{FqtMCT} exhibit the same qualitative 
behavior. With increasing density 
the relaxation becomes progressively slower and the 
fluid approaches a glass transition qualitatively similar
to the one observed with hard particles. A two-step decay 
of time correlation functions emerges, with the slow 
relaxation being strongly dependent on density. 
For the temperatures shown in Fig.~\ref{FqtMCT}, $F(q,t)$
does not decay to zero above $\rho \approx 1.9$ within MCT at $T=0.00287$, 
and above $\rho \approx 2$ within MD at $T=0.003$.
Although the relaxation time is infinite within theory, 
we can simply report that it has become too large to be measured 
in the simulations and that, for practical purposes, the system is a 
non-ergodic glass. In the glass phase there is, at first, 
little change when $\rho$ increases, apart from a weak variation
of the long-time limit of the correlators.
However, quite surprisingly, the glass suddenly melts when 
it is compressed further, and the relaxation time becomes finite again.
This is counter-intuitive from the standpoint 
of glass physics, although perhaps not totally unexpected for a system
known to displaying other anomalies. Past the glass melting 
transition, the dynamics 
becomes faster when density is increased even further. 

\begin{figure}
\includegraphics[width=8.5cm]{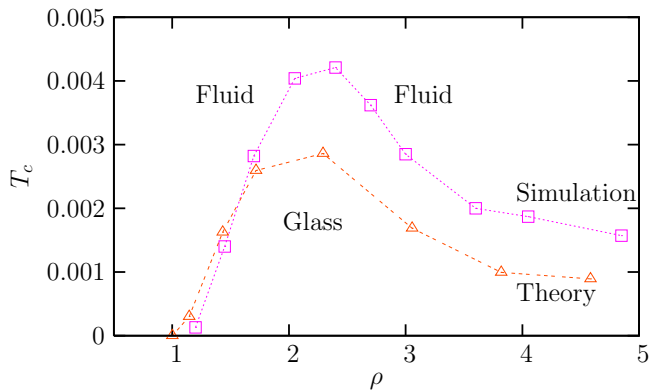}
\caption{\label{phase} Dynamic phase diagram showing the exact 
(theory) and fitted (simulations) critical temperatures 
obtained from power law fits to the structural relaxation times.
A reentrant glass transition is obtained in both cases with a maximum 
near $\rho \approx 2.4$.}
\end{figure}

The above description of time-dependent correlators suggests the existence
of a reentrant glass transition line in the $(\rho,T)$ phase 
diagram, which we confirm 
in Fig.~\ref{phase}. In the theory we easily obtain 
the critical mode-coupling temperature, $T_c(\rho)$, with arbitrary 
precision, by fitting the temperature evolution at constant density of 
the structural relaxation time to a power law divergence, 
$\tau \propto (T-T_c)^{-\gamma}$ \cite{gotze}. 
A finite mode-coupling temperature emerges 
in a way qualitatively similar to the results obtained for the harmonic 
potential~\cite{harmonic}. However, the mode-coupling transition line 
has a clear maximum, $T_c \approx 0.003$, near $\rho \approx 2.4$,
slightly below the density where $S(q)$ has its maximum. It then decreases
smoothly with increasing $\rho$. 
When performed for simulation data, the same analysis works well 
over a moderate range of 2-3 decades of $\tau$ but, 
as usual~\cite{gotze}, the MCT power law breaks down below 
some $T$. Remarkably, however, 
the numerical $T_c$ values reported in Fig.~\ref{phase}.
exhibit trends in good qualitative agreement with theory. 
The important point is the robustness of the reentrant glass transition
line with a maximum around  $\rho\approx 2.4$, at nearly the 
same density where the freezing transition temperature peaks 
for the monodisperse system~\cite{frenkel},
$T_m \approx 0.008$. 

The quantitative disagreement between the two glass lines
has multiple origins. First, the theoretical results were 
obtained for the monodisperse system, but 
the simulations were performed using a polydisperse one. 
Second, both HNC and MCT have their well known limitations.
It is possible to improve the static approximation~\cite{hansen},
but the predicted (spurious) divergence of the relaxation time
predicted by MCT will remain.  
Moreover, MCT predictions stem from the knowledge of pair
correlation functions, incorrectly ignoring additional structural
information~\cite{tarjus}. Nevertheless, in the present case,
the evolution of pair correlations are certainly the major origin of 
the reentrant glass transition, which is thus successfully 
captured by MCT. Analogous successes of MCT 
occurred in the past, for instance for sticky colloidal
particles~\cite{attractivecomment}. 
Despite its limitations, 
our procedure should therefore be useful in revealing
new dynamical features relevant in the context of colloidal 
physics with soft interactions, as our numerical simulations 
confirm.

In summary, by combining simple analytical approaches
and simulations, we predicted and observed numerically
the existence of a reentrant
fluid-glass transition in a system of Hertzian spheres at large densities.
We believe that our results are not restricted to this particular 
interaction and similar behavior should arise in other
systems with soft repulsive potentials, and 
should find experimental realizations using ultrasoft colloids.
Our results suggest that making softer colloids
not only affects the kinetic glass fragility~\cite{tom,dave,ken}, 
but can also produce new, non-intuitive effects such as 
a reentrant glass transition, 
which remains to be seen in an experiment. 

\acknowledgments
We acknowledge financial support from 
ANR Dynhet and R\'egion Languedoc-Roussillon (L. B.), 
MAT2007-63681 and IT-436-07, Spain (A. J. M.), 
and NSF Grant No.\ CHE 0909676 (G. S.).

\end{document}